# Blume-Capel model on cylindrical Ising nanowire with core/shell structure: Existence of a dynamic compensation temperatures

**Mehmet Ertaş and Ersin Kantar**[*]

*Physics Department, Erciyes University, 38039 Kayseri–Turkey*

We present a study, within a mean-field approach, of the kinetics of the spin-1 Blume-Capel model on cylindrical Ising nanowire in the presence of a time-dependent oscillating external magnetic field. We employ the Glauber transition rates to construct the mean-field dynamical equations. We investigate the thermal behavior of the dynamic order parameters. From these studies, we obtain the dynamic phase transition (DPT) points. Then, we study the temperature dependence of the dynamic total magnetization to find the dynamic compensation points as well as to determine the type of behavior. We also investigate the effect of a crystal-field interaction and the exchange couplings between the nearest-neighbor pairs of spins on the compensation phenomenon and construct the phase diagrams in four different planes. The dynamic phase diagrams contain paramagnetic (P), ferromagnetic (F), the antiferromagnetic (AF), and two coexistence or mixed phase regions, namely, the F + P and AF + P that strongly depend on interaction parameters. The system also exhibits the compensation temperatures, or the N-, P-, Q-, S- type behaviors. Furthermore, we also observed two compensation temperatures, namely W-type behaviors, which this result is compared with some experimental works and a good overall agreement is found.

PACS number(s): 05.50. +q, 05.70.Fh, 64.60.Ht, 75.10.Hk

## I. INTRODUCTION

Most magnetic materials exhibit monotonic increases in magnetization with decreasing temperature below their critical temperatures ($T_C$). However, in the classical theory, Néel had envisaged the possibility that the spontaneous magnetization can change sign; i.e., the magnetic poles can invert, at a particular temperature (the so-called compensation temperature, $T_{Comp}$) [1]. Recently, the compensation temperatures has attracted much attention both theoretically [2] and experimentally such as $NiFe_{2-x}V_xO_4$ and $NBu_4[Fe^{II}Fe^{III}(C_2O_4)_3]$ [3]. The reason is that the existence of compensation temperatures is of great technological importance, since at this point only a small driving field is required to change the sign of the total magnetization. This property is very useful in thermomagnetic recording, electronic, and computer technologies [4]. Moreover, using different theoretically models, one-, two- and three-compensation temperatures have been obtained [5]. On the other hand, one observed experimentally two-compensation temperatures in the series $(Ni_{0.22}^{II}Mn_{0.6}^{II}Fe_{0.18}^{II})_{1.5}$ $[Cr^{III}(CN)_6]\cdot7.6H_2O$ with the four sublattices [6] and in the series $K_{0.18}^{I}$ $(Co_{0.39}^{II}Mn_{0.61}^{II})_{1.41}$ $[Cr^{III}(CN)_6]\cdot9H_2O$ with the three sublattices [7].

---

[*] Corresponding author.
 Tel: + 90 (352) 2076666#33136; Fax: + 90 (352) 4374931
 E-mail address: ersinkantar@erciyes.edu.tr (E. Kantar)

On the other hand, magnetic nanoparticle-based systems as for instance nanotube or nanowire have been attracting the attention not only because of their fundamental importance, but also due to many technological applications, such as magnetic fluids, magnetic recording media, spin electronics, optics, sensors, magnetic refrigeration and thermoelectronics devices, among others [8]. The magnetic properties of fine magnetic particles are dominated by a resultant of competition among surface effects, finite size effects and interaction effects. Recently, the Ising model has been applied successfully to the investigation of a nanostructure materials, such as nanowire and nanotube. The equilibrium behaviors of these systems such as hysteresis loops, the compensation behavior, phase transitions phase diagrams, susceptibility, internal energy, specific heat, free energy, reentrant phenomena etc., have been studied by utilizing the effective field theory (EFT), mean field approximation (MFA) and Monte-Carlo (MC) simulation in detail (see [9-11] and references therein).

Although a great amount the Ising systems have used to investigate the equilibrium properties of magnetic nanostructured materials, there have been only a few works that the Ising systems used to investigate dynamic magnetic properties of nanostructured materials [12, 13]. In series of these works, the dynamic behavior of the spin-1/2 Ising model on nanostructured materials were investigated by using the effective-field theory based on the Glauber-type stochastic (DEFT). We should also mention that the dynamic behavior of nanostructure with high spin systems by using the DEFT due to difficulties in implementation to high spin systems of the method were not investigated. Therefore, the dynamic behavior of nanostructure with high spin systems were not studied by utilizing any method. To this end, in present paper, we used to DMFT for study the magnetic properties of a nonequilibrium spin-1 Blume-Capel (BC) cylindrical Ising nanowire system with core/shell in an oscillating magnetic field. The aim of the present paper is three-fold: (i) to obtain the dynamic compensation temperatures and investigate the type of the compensation behavior of the system. (ii) The dynamic phase transitions (DPT) of the Ising nanowire system. (iii) Finally, to present the dynamic phase diagrams, including dynamic compensation temperature, of the system in the different planes.

It is well-known that the DPT in nonequilibrium systems in the presence of an oscillating external magnetic field has attracted much attention in the past few decades, theoretically (see [12-17] and references therein) and analytically [18-20], and experimental evidences for the DPT have been reported in amorphous ultrathin Co films, polyethylene naphthalate (PEN) nanocomposites, cuprate superconductors, [Co/Pt]3 magnetic multilayers, etc. [21-24].

The outline of the paper is as follows. In Sec. II, the model and methodology. The numerical results and discussions are given in Sect. III. A summary and discussion of the results are given in Sect. IV.

## II. MODEL AND METHODOLOGY

The considered model is a spin-1 cylindrical Ising nanowire under the oscillating magnetic field. The schematic representation of a cylindrical Ising nanowire is depicted in Fig. 1, in which the wire consists of the surface shell and the core. Each site on the figure is occupied by a spin-1 Ising particle and each spin is connected to the two nearest-neighbor spins on the above and below sections along the cylinder. The Hamiltonian of the system is given by

$$\mathcal{H} = -J_C\left(\sum_{ii'}\sigma_i\sigma_{i'} + \sum_{ij}\sigma_i S_j + \sum_{jj'}S_j S_{j'}\right) - J_{Int}\left(\sum_{jk}S_j\alpha_k + \sum_{jl}S_j\lambda_l\right)$$
$$-J_S\left(\sum_{kk'}\alpha_k\alpha_{k'} + \sum_{kl}\alpha_k\lambda_l + \sum_{ll'}\lambda_l\lambda_{l'}\right) - H\left(\sum_i\sigma_i + \sum_j S_j + \sum_k \alpha_k + \sum_l \lambda_l\right) \quad (1)$$
$$-D\left(\sum_i(\sigma_i)^2 + \sum_j(S_j)^2 + \sum_k(\alpha_k)^2 + \sum_l(\lambda_l)^2\right),$$

where the $J_S$ and $J_S$ are the exchange interaction parameters between two nearest-neighbor magnetic particles at the surface shell and core, respectively, and $J_1$ is the interaction parameters between two nearest-neighbor magnetic particles at the surface shell and the core shell. D is crystal field. The surface exchange and interfacial coupling interactions are often defined as $J_S = J_C(1+\Delta_S)$ and $r = J_1/J_C$ in the nanosystems [10-12], respectively. H is the oscillating magnetic field: $H(t) = H_0\cos(wt)$, with H0 and w = 2πν being the amplitude and the angular frequency of the oscillating field, respectively. The system is in contact with an isothermal heat bath at an absolute temperature T$_A$.

Now, we apply the Glauber-type stochastic dynamics to obtain the set of the mean-field dynamic equations. Since the derivation of the mean-field dynamic equations was described in detail for spin-1 system [25] and different spin systems [14], in here, we shall only give a brief summary. If the S, α and λ spins momentarily fixed, the master equation for σ- spins can be written as

$$\frac{d}{dt}P^\sigma(\sigma_1,\sigma_2,...,\sigma_N;t) = -\sum_i \left( \sum_{\sigma_i'\sigma_i} W_i^\sigma(\sigma_i \to \sigma_i') \right) P^\sigma(\sigma_1,\sigma_2,...,\sigma_i,...\sigma_N;t)$$
$$+\sum_i \left( \sum_{\sigma_i'\sigma_i} W_i^\sigma(\sigma_i' \to \sigma_i) \right) P^\sigma(\sigma_1,\sigma_2,...,\sigma_i',...\sigma_N;t), \quad (2)$$

where $W_i^\sigma(\sigma_i \to \sigma_i')$ is the probability per unit time that the *i*th spin changes from the value $\sigma_i$ to $\sigma_i'$. Since the system is in contact with a heat bath at absolute temperature $T_A$, each spin can change from the value $\sigma_i$ to $\sigma_i'$ with the probability per unit time;

$$W_i^\sigma(\sigma_i \to \sigma_i') = \frac{1}{\tau} \frac{\exp(-\beta \Delta E^\sigma(\sigma_i \to \sigma_i'))}{\sum_{\sigma_i'} \exp(-\beta \Delta E^\sigma(\sigma_i \to \sigma_i'))}, \quad (3)$$

where $\beta = 1/k_B T_A$, $k_B$ is the Boltzmann factor, $\sum_{\sigma_i'}$ is the sum over the three possible values of $\sigma_i' = \pm 1, 0$, and

$$\Delta E^\sigma(\sigma_i \to \sigma_i') = -(\sigma_i' - \sigma_i)\left( H + z_{\sigma\sigma} J_C \sum_{i'} \sigma_{i'} + z_{\sigma S} J_C \sum_j S_j \right) - \left[ (\sigma_i')^2 - (\sigma_i)^2 \right] D, \quad (4)$$

gives the change in the energy of the system when the $\sigma_i$-spin changes. The probabilities satisfy the detailed balance condition. Using Eqs. (2), (3), (4) with the mean-field approach, we obtain the mean-field dynamic equation for the σ-spins as

$$\Omega \frac{d}{d\xi} m_{C1} = -m_{C1} + \frac{2\sinh[\beta(x_1 + H_0 \cos(\xi))]}{2\cosh[\beta(x_1 + H_0 \cos(\xi))] + \exp[-d/T]}, \quad (5)$$

where $x_1 = z_{\sigma\sigma} J_C m_{C1} + z_{\sigma S} J_C m_{C2}$, $m_{C1} = \langle \sigma \rangle$, $m_{C2} = \langle S \rangle$, $\xi = wt$ and $\Omega = \tau w = w/f$, w is the frequency of the oscillating magnetic field and f represents the frequency of spin flipping. Moreover, $z_{\sigma\sigma}$ and $z_{\sigma S}$ corresponds to the number of nearest-neighbor pairs of spins σ–σ and σ–S, respectively, in which $z_{\sigma\sigma} = 2$ and $z_{\sigma S} = 6$. As similar to σ-spins, we obtain the mean-field dynamical equations for the S, α and λ -spins by using the similar calculations. The mean-field dynamic equations for S, α and λ -spins are obtained as

$$\Omega \frac{d}{d\xi} m_{C2} = -m_{C2} + \frac{2\sinh[\beta(x_2 + h_0 \cos(\xi))]}{2\cosh[\beta(x_2 + h_0 \cos(\xi))] + \exp[-d/T]}, \tag{6}$$

$$\Omega \frac{d}{d\xi} m_{S1} = -m_{S1} + \frac{2\sinh[\beta(x_3 + h_0 \cos(\xi))]}{2\cosh[\beta(x_3 + h_0 \cos(\xi))] + \exp[-d/T]}, \tag{7}$$

$$\Omega \frac{d}{d\xi} m_{S2} = -m_{S2} + \frac{2\sinh[\beta(x_4 + h_0 \cos(\xi))]}{2\cosh[\beta(x_4 + h_0 \cos(\xi))] + \exp[-d/T]}, \tag{8}$$

where $h_0 = H_0/J_C$, $d = D/J_C$, $x_2 = z_{S\sigma}J_C m_{C1} + z_{SS}J_C m_{C2} + z_{S\alpha}J_1 m_{S1} + z_{s\lambda}J_1 m_{S2}$, $x_3 = z_{\alpha\alpha}J_S m_{S1} + z_{\alpha\lambda}J_S m_{S2} + z_{\alpha S}J_1 m_{C2}$, $x_4 = z_{\lambda\lambda}J_S m_{S2} + z_{\lambda\alpha}J_S m_{S1} + z_{\lambda S}J_1 m_{C2}$, $m_{S1} = \langle\alpha\rangle$, $m_{S2} = \langle\lambda\rangle$, $z_{S\sigma} = z_{S\alpha} = z_{\alpha S} = 1.0$, $z_{SS} = 4$, $z_{\alpha\lambda} = z_{\lambda\lambda} = z_{\lambda\alpha} = z_{\lambda S} = 2.0$. Hence, a set of mean-field dynamical equations of the system are obtained. We fixed $J_C = 1.0$ that the core shell interaction is ferromagnetic and $\Omega = 2\pi$.

### III. NUMERICAL RESULTS AND DISCUSSION

#### A- Phases in the system

In order to find phases in the system, we shall present the stationary solutions of the set of DMFT equations, namely Eqs. (5)-(8). The stationary solutions of these equations will be periodic functions of $\xi$ with period $2\pi$; that is, $m_{S1,S2}(\xi+2\pi) = m_{S1,S2}(\xi)$ and $m_{C1,C2}(\xi+2\pi) = m_{C1,C2}(\xi)$. Furthermore, they can be one of the two types according to whether they have or do not have the properties

$$m_{S1,S2}(\xi+2\pi) = -m_{S1,S2}(\xi), \tag{9a}$$

and

$$m_{C1,C2}(\xi+2\pi) = -m_{C1,C2}(\xi). \tag{9b}$$

The first type of solution satisfies Eqs. (9a) and (9b) which is called a symmetric solution which corresponds to a paramagnetic solution or phase. In this solution, the shell and core magnetizations $m_{S1,S2}$ and $m_{C1,C2}$ are equal to each other ($m_{S1,S2} = m_{C1,C2}$); and they oscillate around zero and are delayed with respect to the external magnetic field. The second type of solution, which does not satisfy Eqs. (9a) and (9b), is called a nonsymmetric solution that corresponds to a ferromagnetic solution or phase. In this solution, the shell and core magnetizations are not equal each other ($m_{S1,S2} = m_{C1,C2} \neq 0$); and they oscillate around a positive value. In this case, the shell and core magnetizations do not follow the external magnetic field. Hence, if $m_{C1,C2}(\xi)$, $m_{S1,S2}(\xi)$ oscillate around +1, the solution is called the ferromagnetic (F) solution or phase. If $m_{C1,C2}(\xi)$ oscillates around ±1, $m_{S1,S2}(\xi)$ oscillates around ∓1, the solution is called the antiferromagnetic phase (AF). We found that a P, F, AF fundamental phases in this system. Moreover we also obtain two mixed phases, namely the F + P in which F, P phases coexist and the AF + P in which AF, P phases coexist. Since we gave the solution of these kinds of dynamic equations in Ref. 12, 15 in detail, we will not discuss the solutions and present any figures here.

**B- Thermal behavior of dynamic core/shell magnetizations and the total magnetization**
*B.1-The Dynamic Phase Transitions (First- and Second-Order)*

We have to calculate DPT points in order to see the dynamic phase boundaries among the obtained phase regions. To this end, we solved numerically Eqs. (5)-(8) by means of the Adams–Moulton predictor corrector method and Romberg integration method. Consequently, the behaviors of the dynamic core/shell magnetizations as a function of the temperature are examined for different values of the interaction parameters. Also, this examination leads us to characterize the nature (first- or second-order) of phase transitions. The dynamic core/shell magnetizations are defined as

$$M_{C1,C2} = \frac{1}{2\pi}\int_0^{2\pi} m_{C1,C2}(\xi)d\xi, \tag{10a}$$

$$M_{S1,S2} = \frac{1}{2\pi}\int_0^{2\pi} m_{S1,S2}(\xi)d\xi, \tag{10b}$$

To obtain the DPT points as well as to characterize the nature of the dynamic boundaries, the behaviors of $M_{C1,C2}$ and $M_{S1,S2}$ as functions of the temperature for several values of interaction parameters are obtained by solving Eqs. (10a)-(10b) combined the numerical methods of the Adams-Moulton predictor corrector with Romberg integration. A few interesting results are plotted in Figs. 2(a)-(d). While Figs. 2(a) and (b) show a second-order phase transition, Fig. 2(c) and (d) illustrates a first-order phase transition. In these figures, $T_C$ and $T_t$ are the critical or the second-order phase transition and first-order phase transition temperatures. Fig. 2(a) demonstrates the thermal variation of the dynamic core and shell magnetizations for $r=1.0$, $\Delta_S=-0.5$, $d = 0.0$ and $h_0 = 0.1$. In this figure, the dynamic magnetizations 1.0 at zero temperature, and they decreases to zero continuously as the temperature increases; therefore, a second-order phase transition occurs at $T_C = 4.28$ and the phase transition is from the ferromagnetic (F) phase to the paramagnetic (P) phase. Fig. 2(b) shows the temperature dependence of the magnetizations for r =-1.0, $\Delta_S$=-0.5, d = 0.0 and $h_0$= 0.1. In this figure, $M_{C1,C2}$=-1.0 and $M_{S1,S2}$=1.0 at zero temperature. While the dynamic core magnetizations increases, the shell magnetizations decrease continuously with the increasing of the values of temperature below the critical temperature and they become zero at $T_C$ = 4.28; therefore, a second-order phase transition occurs. The transition is from the AF phase to the P phase. Fig. 2(c) illustrates the thermal variations of the dynamic magnetizations for r =1.0, $\Delta_S$=-0.5 d=0.0 and $h_0$=5.0. In Fig. 3(c), the system undergoes a first-order phase transition because discontinuity occurs for the dynamic $M_{C1,C2}$ and $M_{S1,S2}$. The dynamic transition is from the F phase to the P phase at $T_t$ = 1.76. In Fig. 2d, as similar to Fig. 2c, we obtained the system undergoes a first-order phase transition for r =1.0, $\Delta_S$=-0.9 d=0.0 and $h_0$=4.5. In this figure, a first-order phase transition from the i phase to the p phase at $T_t$= 0.70.

### *B.2- The Temperature Dependence of the Total Magnetization and Compensation Types*

The dynamic compensation temperature and the type of compensation behavior are obtained by investigating the behavior of the dynamic total magnetization as a function of the temperature. The dynamic total magnetizations $M_t = (m_{C1} + 6m_{C2} + 6(m_{S1} + m_{S2}))/19$ as the time-averaged magnetization over a period of the oscillating magnetic field are defined as

$$M_t = \frac{1}{2\pi} \int_0^{2\pi} \left( \frac{m_{C1}(\xi) + 6m_{C2}(\xi) + 6(m_{S1}(\xi) + m_{S2}(\xi))}{19} \right) d\xi . \qquad (11)$$

The dynamic compensation temperature, which dynamic total magnetization ($M_t$) vanishes at the compensation temperature $T_{Comp}$. The compensation point can then be determined by looking for the crossing point between the absolute values of the surface and the core magnetizations. Therefore, at the compensation point, we must have

$$\left| M_{Surface}(T_{Comp}) \right| = \left| M_{Core}(T_{Comp}) \right|, \qquad (12)$$

and

$$\text{sgn}\left[ M_{Surface}(T_{Comp}) \right] = -\text{sgn}\left[ M_{Core}(T_{Comp}) \right]. \qquad (13)$$

We also require that $T_{Comp} < T_C$, where $T_C$ is the critical point temperature.

Consequently, we obtained the N-, P-, Q-, S-type behaviors as well as W-type two compensation behaviors in the Ising nanowire system. Fig. 3 shows the temperature dependencies of the total magnetization $M_t$ for several values of r, $\Delta_S$, D and $h_0$. As seen from Fig. 3(a), the curve labeled r = -1.0, $\Delta_S$ =-0.5, d = 0.0 and $h_0$ = 0.1 may show the N-type behavior. Moreover, the P-type behavior is obtained in Fig. 3(b) for r = -0.1, $\Delta_S$ =1.0, d = 0.0 and $h_0$ = 0.1. Fig. 3(c) is calculated for r = -1.0, $\Delta_S$ =0.5, d = 0.0 and $h_0$ = 0.1. As we can see this curve illustrates the Q- type behavior. Fig. 3(d) is calculated for r = 0.1, $\Delta_S$ =-0.9, d = 0.0 and $h_0$ = 0.1 and illustrates the S-type behavior. Finally, we obtained W-type two compensation behavior for r = -1.0, $\Delta_S$ =0.0, d = 0.0 and $h_0$ = 5.0, seen in Fig. 3(e). We should mention that the very similar behavior with Fig. 3(e) were also obtained in experimental magnetization versus temperature curves for $(Ni^{II}_{0.22}Mn^{II}_{0.60}Fe^{II}_{0.18})_{1.5}[Cr^{III}(CN)_6]_{7.6}H_2O$, which exhibits magnetization reversals at 35 and 53 K [6]. In more clear words, this behavior similar to Fig. 3 of the Ref. 6. At this point, we should also mentioned that two compensation points of a mixed ferro–ferrimagnetic ternary alloy of the type $AB_pC_{1-p}$ with different spin $S_{Ni}$=1, $S_{Mn}$=5/2 and $S_{Cr}$=3/2 by Monte Carlo simulations [26], $(A_aB_bC_c)_yD)$ with different spin $S_{Ni}$=1, $S_{Mn}$=5/2, $S_{Fe}$=2, $S_{Cr}$=3/2 by the mean-field theory [27] and ternary system with a single-ion anisotropy on the Bethe lattice [28] have been found. Furthermore, experimentally [6, 7], two-compensation points have been also observed.

### C- Dynamic phase diagrams

Since we have examined the nature of the DPT and dynamic compensation points in Subsection III.A and Subsection III.B, we can now obtain the dynamic phase diagrams of the system. In Figs. (4)-(5), the dynamic phase diagrams in the four different planes, namely (h, T), (d, T), (r, T) and ($\Delta_S$, T) are presented for various values of interaction parameters. In these phase diagrams, the solid lines, dashed lines and dashed-dot lines represent the second- and first-order phase transition lines, and compensation temperature lines respectively, and the dynamic tricritical points is denoted by a filled circle.

Fig. 4 displays the dynamic phase diagrams in the (d, T) plane for positive (ferromagnetic cases, r > 0) and negative (antiferromagnetic cases, r < 0) values of interfacial coupling and six main different topological types of phase diagrams are seen. From these phase diagrams following interesting phenomena have been observed. The phase diagrams of Figs. 4(a), (c), (d) and (f), illustrate tricritical points, but Figs. 4(b) and 4(e) do not contain any tricritical point. Moreover, in Figs. 4(d)-(f), the phase diagrams exhibits compensation points.

In Fig. 4(e), the system also exhibits two compensation point behavior for high values of crystal field.

We also calculate the dynamic phase diagrams in the ($h_0$, T), (r, T) and ($\Delta_S$, T) planes, seen in Fig. 5, respectively. If one can investigate these phase diagrams, one can observe to contain the compensation temperatures of Figs. 5(a)-(c). Fig. 5(a) obtained for values of $r$ = -0.5, $\Delta_S$ = -0.9 and d = 0.0. Fig. 5(a) illustrates the dynamic phase diagram in the ($h_0$, T) plane and contains dynamic tricritical points and the compensation temperature for low values of $h_0$ and T. Fig. 5(b) displays the dynamic phase diagram in the (r, T) plane and plotted for the values of $h_0$ = 0.1, $\Delta_S$ = -0.5 and d = -1.5. In Fig. 5(b), we can clearly see that phase transitions are the second-order and first- order for high and low values of T, respectively. In the ($\Delta_S$, T) plane, dynamic phase diagram presented in Fig. 5(c). This figure is obtained for values of $h_0$ = 0.1, $J_1$ = -1.0 and d = -1.5.

## IV. SUMMARY AND CONCLUSION

We have analyzed, within a mean-field approach, the stationary states kinetics of the Blume-Capel model on cylindrical Ising nanowire under the presence of a time varying (sinusoidal) magnetic field (H = $H_0$ cos(wt)). We use a Glauber-type stochastic dynamics to describe the time evolution of the system. First we have studied time variations of the average order parameters in order to determine the phases in the systems. Then, we have investigated the dynamic order parameters as a function of the temperature to characterize the nature (continuous and discontinuous) of transitions as well as to find the DPT points and compensation points. Lastly, the dynamic phase diagrams are presented in the ($h_0$, T), (d, T), (r, T) and ($\Delta_S$, T) planes. We have found six different dynamic phase diagrams topologies in the (d, T) plane. We observed the system undergoes first- and second-order phase transitions, tricritical point as well as the compensation points for the certain values of Hamiltonian parameters. Moreover, N-, P-, Q-, S-and W- types of compensation behaviors exist in the system that are also strongly dependent on interaction parameters; hence, one or two compensation points have been found. Two compensation behavior, namely W-type, is compared with some experimental works and a good overall agreement is found.

Finally, up to this time, the dynamic behaviors of nanostructure (nanoparticle, nanowire, nanotube etc.) were only studied for spin-1/2 Ising system by using an effective-field theory based on the Glauber-type stochastic dynamics (DEFT). Unfortunately, the dynamic behavior of nanostructure with high spin systems by using the DEFT due to difficulties in implementation to high spin systems of the method were not investigated. Therefore, we hope that the presented paper may be pioneer for researchers who want to examine dynamic behavior of high spin nanostructure. Moreover, we also hope this study will contribute to the theoretical and experimental research on the dynamic magnetic properties of nanostructure Ising systems as well as to research on magnetism.

### LIST OF THE FIGURE CAPTIONS

**FIG. 1. (color online)** Schematic representations of a cylindrical nanowire: **(a)** cross-section and **(b)** three-dimensional. The gray and blue circles indicate spin-1 Ising particles at the surface shell and core, respectively. (For interpretation of the references to color in this figure legend, the reader is referred to the web version of this article).

**FIG. 2. (color online)** The temperature dependence of the dynamic core and shell magnetizations. $T_C$ and $T_t$ are the second-order and first-order phase transition temperatures, respectively.

- **(a)** Exhibiting a second-order phase transition from the F phase to P phase at $T_C$ = 4.28 for r=1.0, $\Delta_S=-0.5$, d=0.0 and $h_0$ = 0.1.

- **(b)** Exhibiting a second-order phase transition from the F phase to P phase at $T_C$ = 4.28 for r=−1.0, $\Delta_S=-0.5$, d=0.0 and $h_0$ = 0.1.

- **(c)** Exhibiting a first-order phase transition from the F phase to P phase at $T_t$ = 1.76 for r=1.0, $\Delta_S=0.0$, d=0.0 and $h_0$ = 5.0.

- **(d)** Exhibiting a first-order phase transition from the AF phase to P phase at $T_t$ = 0.70 for r=−0.5, $\Delta_S=-0.9$, d=0.0 and $h_0$ = 4.5.

**FIG. 3. (color online)** The dynamic total magnetization as a function of the temperature for different values of interaction parameters. The system exhibits the N-, P-, Q-, S-, W- type

behaviors of compensation behaviors. $T_{comp}$ is the compensation temperature. Dash-dot lines represent the compensation temperatures,

- **(a)** r = -1.0, $\Delta_S$ = -0.5, d = 0.0 and $h_0$ = 0.1,
- **(b)** r = -0.1, $\Delta_S$ = 1.0, d = 0.0 and $h_0$ = 0.1,
- **(c)** r = -1.0, $\Delta_S$ = 0.5, d = 0.0 and $h_0$ = 0.1,
- **(d)** r = 0.1, $\Delta_S$ = -0.9, d = 0.0 and $h_0$ = 0.1,
- **(e)** r = -1.0, $\Delta_S$ = 0.0, d = 0.0 and $h_0$ = 5.0.

**FIG. 4. (color online)** The dynamic phase diagrams for (d, T) plane. Dashed and solid lines are the dynamic first- and second-order phase boundaries, respectively. The dash-dot line illustrates the compensation temperatures. The dynamic tricritical points are indicated with filled circles. **(a)** r=1.0, $\Delta_S$=0.0 and $h_0$=0.1; **(b)** r=1.0, $\Delta_S$=0.0 and $h_0$=4.5; **(c)** r=1.0, $\Delta_S$=0.0 and $h_0$=5.0; **(d)** r=-1.0, $\Delta_S$=-0.5 and $h_0$=0.1; **(e)** r=-1.0, $\Delta_S$=0.0 and $h_0$=4.5; **(f)** r=-1.0, $\Delta_S$=0.0 and $h_0$=5.0;

**FIG. 5. (color online)** Same as Fig 4, but the dynamic phase diagrams for ($h_0$, T), (r, T), ($\Delta_S$, T) planes. **(a)** r=-0.5, $\Delta_S$=-0.9 and d = 0.0; **(b)** $\Delta_S$=-0.5, d=-1.5 and $h_0$= 0.1; **(c)** r=-1.0, d=-1.5 and $h_0$=0.1.

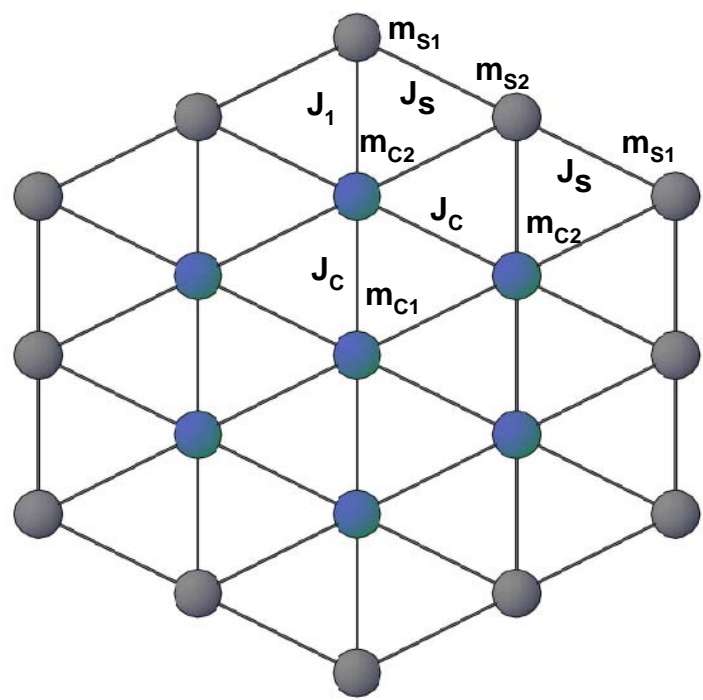

**(a)**

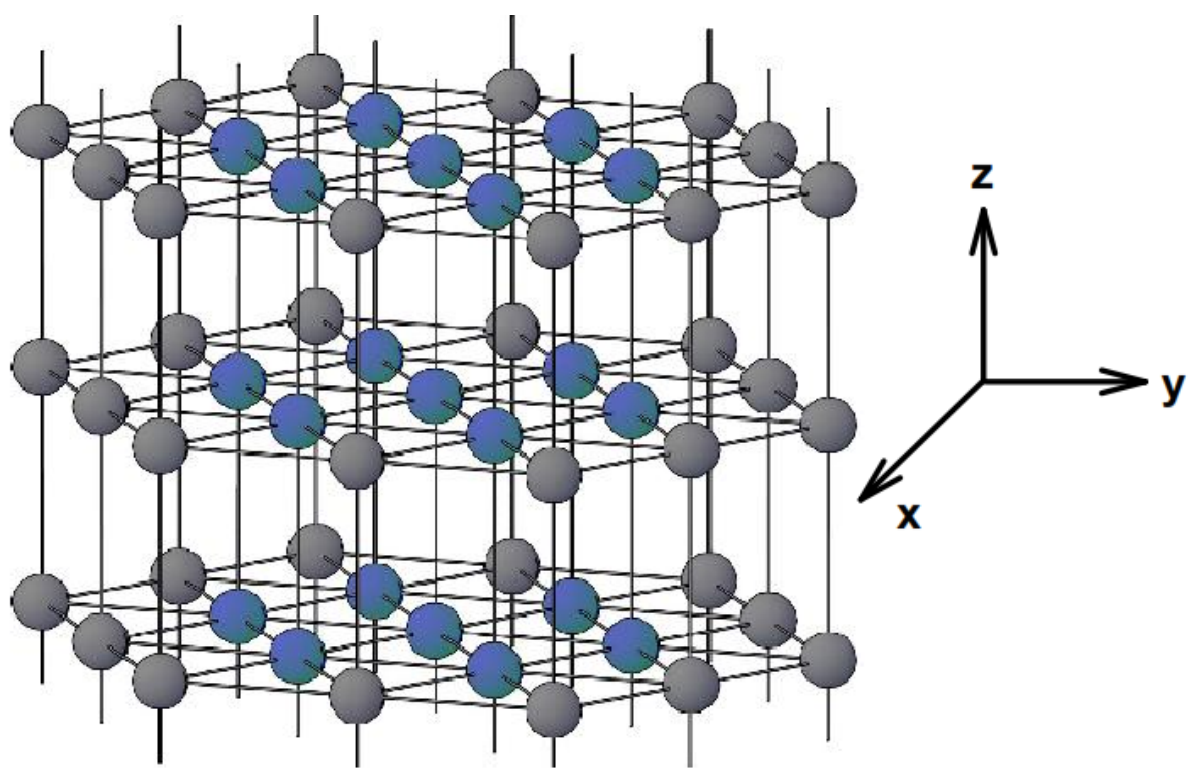

**(b)**

FIG. 1

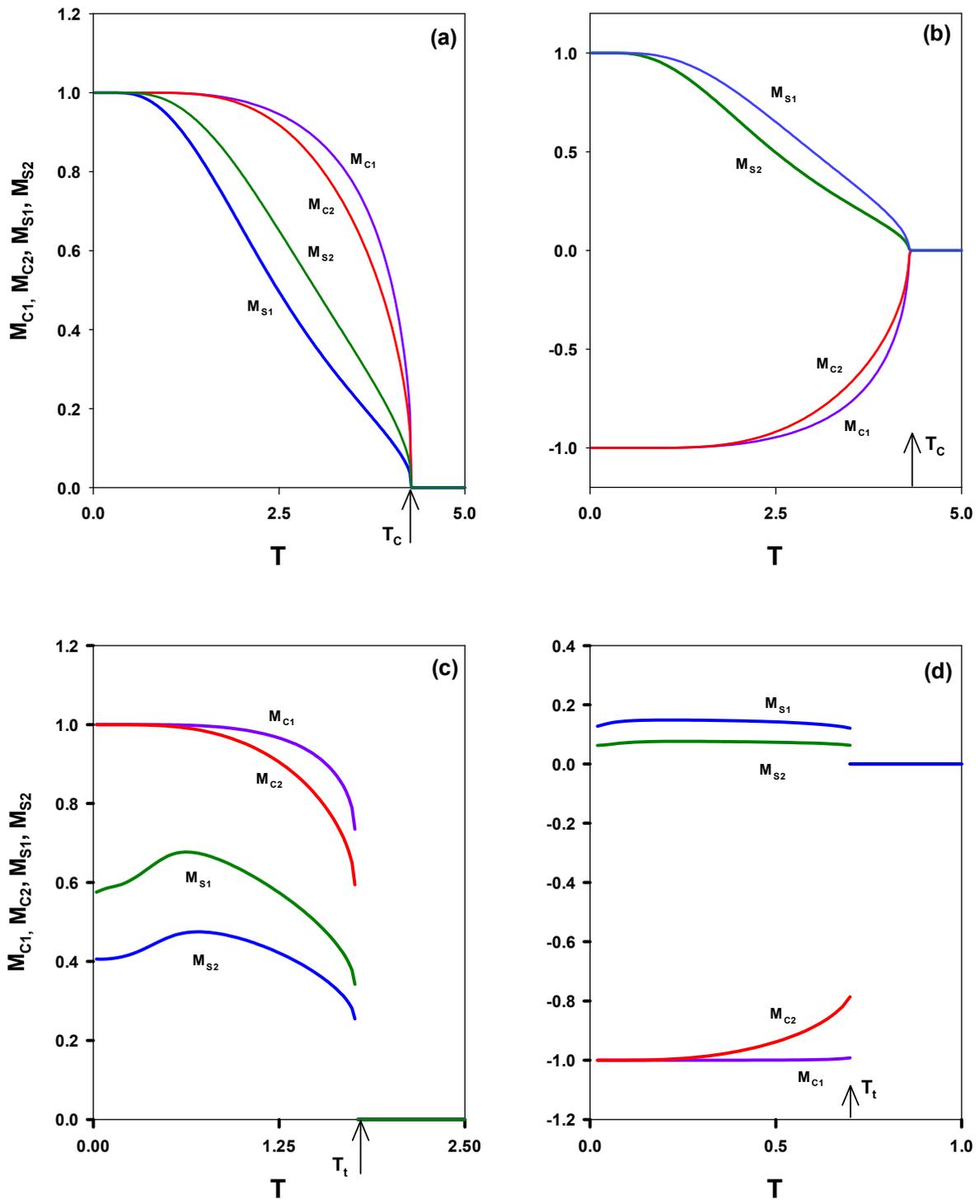

FIG. 2

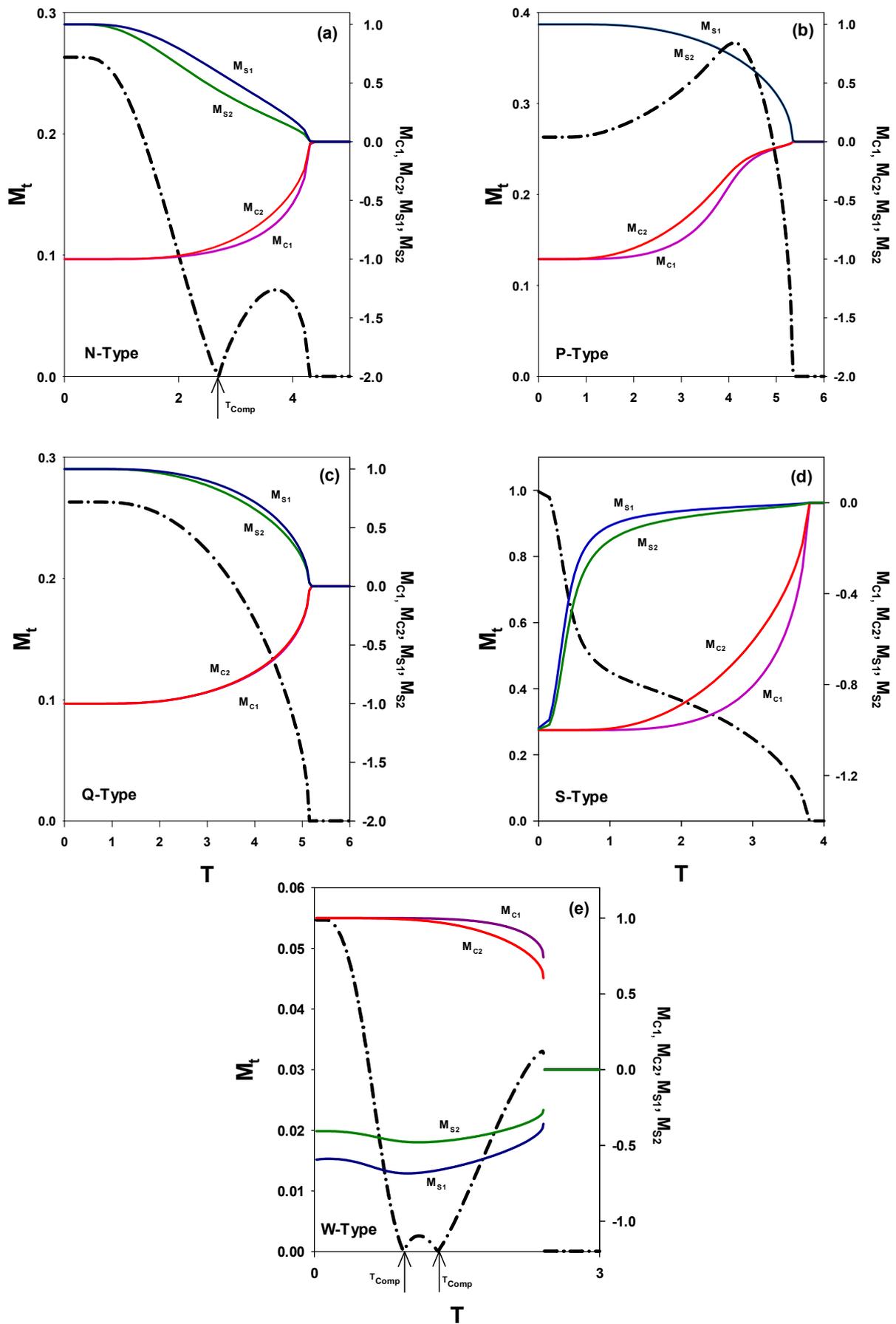

FIG. 3

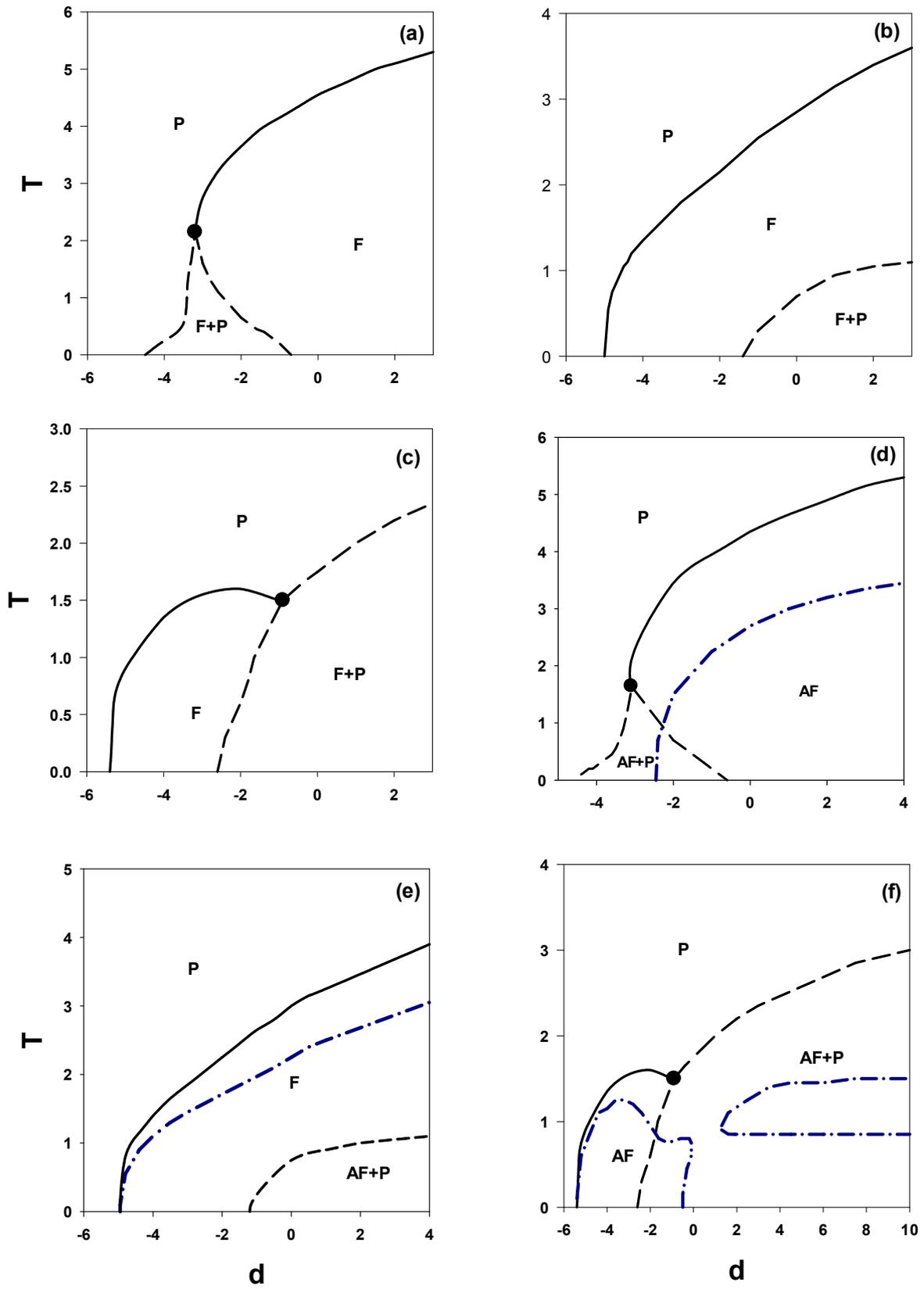

FIG. 4

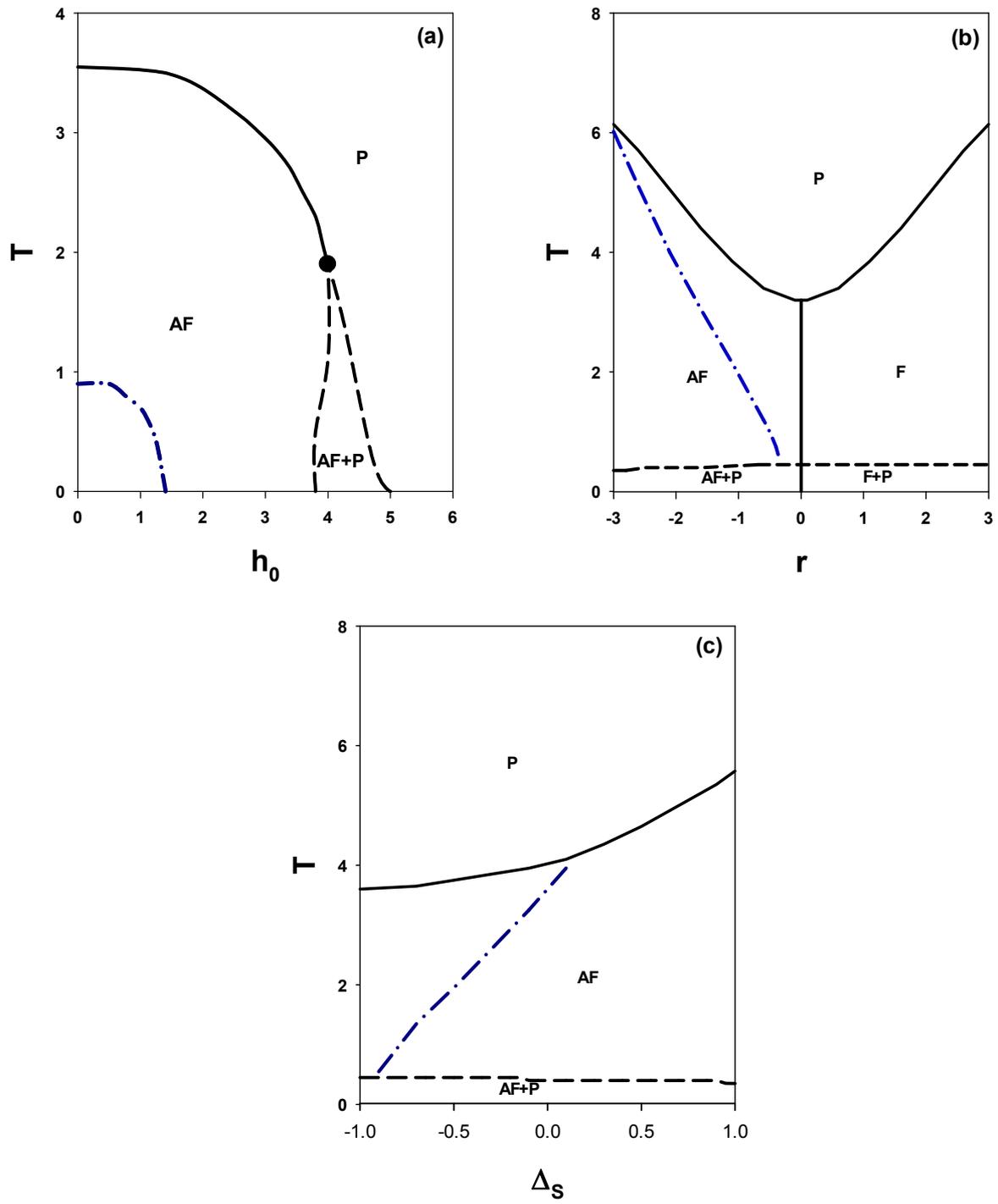

**FIG. 5**